\documentclass[12pt, draftclsnofoot, onecolumn]{IEEEtran}
\usepackage{epsf}
\usepackage{amsmath,amssymb}
\usepackage{graphicx}
\usepackage{color}
\usepackage{cite}
\usepackage{multirow,tabularx} 
\usepackage{ifthen}
\usepackage{times}
\usepackage{stackrel}
\usepackage{afterpage}
\usepackage{float}

\newcommand{\hide}[1]{\ifthenelse{\boolean{false}}{#1}{}}

\newcommand{\qed}{\nobreak \ifvmode \relax \else
      \ifdim\lastskip<1.5em \hskip-\lastskip
      \hskip1.5em plus0em minus0.5em \fi \nobreak
      \vrule height0.75em width0.5em depth0.25em\fi}

\newcommand{\barr}{\begin{array}}
\newcommand{\earr}{\end{array}}

\newcommand{\benum}{\begin{enumerate}}
\newcommand{\eenum}{\end{enumerate}}

\newcommand{\bit}{\begin{itemize}}
\newcommand{\eit}{\end{itemize}}

\newcommand{\bc}{\begin{center}}
\newcommand{\ec}{\end{center}}

\newcommand{\bdes}{\begin{description}}
\newcommand{\edes}{\end{description}}

\newcommand{\bfig}{\begin{figure}}
\newcommand{\efig}{\end{figure}}

\newcommand{\bemq}{\begin{quote} \begin{em}}
\newcommand{\eemq}{\end{em} \end{quote}}

\newcommand{\bmp}{\begin{minipage}}
\newcommand{\emp}{\end{minipage}}

\newcommand{\eqn}[1]{(\ref{#1})}

\newcommand{\brac}[1]{\left({#1}\right)}
\newcommand{\sbrac}[1]{\left[{#1}\right]}
\newcommand{\cbrac}[1]{\left\{{#1}\right\}}

\newcommand{\kth}{^{{\mathrm{th}}}}

\newcommand{\define}{\triangleq}

\newcommand{\ie}{{\it i.e.}}

\newcommand{\iid}{{i.i.d.}}

\newcommand{\bsp}{\begin{slide*}}
\newcommand{\esp}{\end{slide*}}
\newcommand{\bsl}{\begin{slide}}
\newcommand{\esl}{\end{slide}}

\newcommand{\abs}[1]{\left\lvert{#1}\right\lvert}

\IEEEoverridecommandlockouts

\begin{document}

\title{3D MIMO Outdoor-to-Indoor Propagation Channel Measurement}

\author{V. Kristem, S. Sangodoyin, {\it Student Member, IEEE}, \\ C. U. Bas, {\it Student Member, IEEE},  M. K$\ddot{\text{a}}$ske, 
J. Lee, {\it Senior Member, IEEE}, \\ C. Schneider,  G. Sommerkorn, 
J. Zhang, {\it Fellow, IEEE},  \\ R. S. Thom$\ddot{\text{a}}$, {\it Fellow, IEEE}, and A. F. Molisch, {\it Fellow, IEEE} 

\thanks{Part of this work has appeared in Globecom, 2015~\cite{kristem_globecomm}. Part of this work was supported by an NSF MRI grant.}
\thanks{V. Kristem, S. Sangodoyin, C. U. Bas,  A. F. Molisch are with the University of Southern California, Los Angeles, USA. M. K$\ddot{\text{a}}$ske, C. Schneider,  G. Sommerkorn, R. S. Thom$\ddot{\text{a}}$ are with the Technical University Ilmenau, Germany. J. Lee is with Samsung, Korea. J. Zhang is with Samsung USA, Dallas, USA. (Email: \{kristem; sangodoy; cbas; molisch\}@usc.edu; \{Martin.Kaeske; christian.schneider; gerd.sommerkorn; reiner.thomae\}@tu-ilmenau.de; \{juho95.lee;  jianzhong.z\}@samsung.com)}
}
\maketitle\newcommand{\ceil}[1]{\lceil{#1}\rceil}
\vspace{-1.7cm}
\begin{abstract}
3-dimensional Multiple-Input Multiple-Output (3D MIMO) systems have received great interest recently because of the spatial diversity advantage and capability for full-dimensional beamforming, making them promising candidates for practical realization of massive MIMO. In this paper, we present a low-cost test equipment (channel sounder) and post-processing algorithms suitable for investigating 3D MIMO channels, as well as the results from a measurement campaign for obtaining elevation and azimuth characteristics in an outdoor-to-indoor (O2I) environment. Due to limitations in available antenna switches, our channel sounder consists of a hybrid switched/virtual cylindrical array with effectively 480 antenna elements at the base station (BS). The virtual setup  increased the overall MIMO measurement duration, thereby introducing phase drift errors in the measurements. Using a reference antenna measurements, we estimate and correct for the phase errors during post-processing. We provide the elevation and azimuth angular spreads, for the measurements done in an urban macro-cellular (UMa) and urban micro-cellular (UMi) environments, and study their dependence on the UE height. 

Based on the measurements done with UE placed on different floors, we study the feasibility of separating users in the elevation domain. The measured channel impulse responses are also used to study the channel hardening aspects of Massive MIMO and the optimality of Maximum Ratio Combining (MRC) receiver.
\end{abstract}

\vspace{-0.5cm}
\begin{IEEEkeywords}
Outdoor-to-Indoor, Measurement techniques, MIMO, Phase measurement, Channel models
\end{IEEEkeywords}

\section{Introduction}
Multiple-input-multiple-output (MIMO) systems can give tremendous performance improvements over single antenna systems because of their beamforming and spatial multiplexing capabilities~\cite{winters_mimo,Foschini_1998,Molisch2010WirelessBook}. However, the performance gains depend on the propagation channel in which the communication system is deployed. Hence it is important to characterize the channel to get a more accurate assessment of the performance of communication systems. Much of the channel measurements and modeling literature has focused only on the azimuth dispersion~\cite{Almers_2007,3gpp,Meinila}, as the number of antenna elements at each link end is limited and the dispersion in azimuth is much larger than the dispersion in elevation. In terms of scenarios, outdoor-to-outdoor and indoor-to-indoor propagation environments were given most emphasis.

Recently, massive MIMO gained lot of attention because of its ability to improve the spectrum efficiency (SE) and energy efficiency (EE) as well as simplify signal processing, and has been regarded as a promising technique for next generation wireless communication networks~\cite{marzetta_2010,Rusek_2013}. The base station (BS), which may be equipped with hundreds of antenna elements, can be used to serve several tens of users simultaneously. While linear arrays with hundreds of elements pose both scientific and practical challenges, three-dimensional (3D) arrays provide an attractive way of implementation. For example, a cylindrical array structure with a large number of antennas can serve users distributed in elevation and azimuth domains. Under the name of full-dimensional MIMO (FD-MIMO), this approach has been explored in both the scientific literature and international standardization in 3GPP~\cite{Zhang_et_al_2014}. Thus it is important to characterize both elevation and azimuth dispersion of the environments~\cite{molisch_2014}. Furthermore, due to the prevalence of smart phones, most cellular connections are now between an indoor user equipment (UE) and an outdoor base station (BS). Thus, outdoor-to-indoor (O2I) channel characterization for 3D MIMO systems is highly relevant for the fifth generation (5G) cellular systems.

{\em Previous work: }Angular spreads for outdoor environments has been studied in~\cite{Schneider_2010,Laurila_2002,Toeltsch_2002,medbo_2012,Sommerkorn_2014,Pei_VTC_2013_Outdoor,Zhang_TVT_2016}:~\cite{Schneider_2010} measured elevation spread at UE,~\cite{Laurila_2002,Toeltsch_2002,medbo_2012} measured elevation spread at BS,~\cite{Sommerkorn_2014,Pei_VTC_2013_Outdoor,Zhang_TVT_2016} measured the elevation and azimuth spreads at the BS and UE. Angular spreads for indoor environments has been studied in~\cite{Quitin_2010,Huang_2009,Zhao_IET_2016_D2D,Zhang_ICC_2014_indoor}:~\cite{Quitin_2010} measured elevation spread at UE,~\cite{Huang_2009,Zhao_IET_2016_D2D,Zhang_ICC_2014_indoor} measured elevation and azimuth spreads at the BS and UE. Ref.~\cite{molisch_2014} and~\cite{Molisch2014BookCh} are the survey publications describing the 3D MIMO outdoor and indoor channel measurements and models. Ref.~\cite{wyne_2008,Oestges_2010} measured the azimuth spreads for an O2I environment, using linear antenna arrays. There are very few papers that study the elevation characteristics of O2I environments. Ref.~\cite{Rui_2014,Wang_2014,Kavya_MILCOM_2013} studies the elevation spreads for an O2I environment, and models the dependency of the elevation spread on the UE height, using the ray tracing simulations. Ref.~\cite{Omaki_2014} measured the elevation spreads at BS, using a cylindrical antenna array at the BS and a single antenna at the UE, for an O2I micro-cellular environment. Ref.~\cite{Zhang_TVT_2016,Zhang_2015_ICNC} and Ref.~\cite{Zhang_2015_WCNC} measured the elevation spreads at the BS and UE for a macro and micro cellular environments respectively. These works used a 16 element dual-polarized uniform planar antenna array (4x4 matrix) at the BS. More such measurements are required to better characterize the angular spreads at the BS and UE. In the current paper, we present the results of an O2I 3D MIMO channel measurement campaign carried out in an urban macro-cellular (UMa) and urban micro-cellular (UMi) environments using a 480-element antenna array at the BS. The large antenna array at the BS, enables us to better resolve the multipath components (MPCs) at the BS. Preliminary versions of our results were provided to the 3GPP standardization group for FD-MIMO, and were taken into account for the establishment of their channel model~\cite{3gpp_3DMIMO}.

One of the key challenges of the 3D MIMO O2I channel measurements is the complexity and the cost of the transmitter setup. Typically, MIMO channel measurements are obtained using the switched array principle, where a single transmit/receive radio frequency (RF) chain is successively connected to elements of a transmit/receive antenna array~\cite{Thoma_2001,Thoma_2000}. Since the BS is equipped with several hundreds of antenna elements and the path loss of O2I environment is relatively large, this demands for an  electronic switch with large number of ports and low insertion loss, which translates to increased cost. 

We propose an alternative sounding technique where a virtual cylindrical array is created using a vertical linear array and a mechanical rotor (to scan the azimuth plane). While this setup only requires a switch with enough ports to switch between the elements of the linear array, the MIMO measurement duration significantly increases. The non-idealities of the clocks at the transmitter and receiver introduce phase errors in the channel transfer function measurements that increase over time. Even highly precise clocks, as used in our case, lead to significant errors. For short distance measurements, the transmit and receive clocks can be synchronized using fiber or electrical cables. But, for the O2I measurements with transmitter-receiver separation of over 150~m, it is difficult to run the cables from the transmitter to receiver because of the moving traffic on the roads or pedestrians. Hence, using {\em over-the-air} clock synchronization approach where a reference antenna measurement can be used to track the phase drift in the clocks over time, and is further used to correct for the phase errors during the post processing.
 
The key contributions of the paper are as follows
\bit
\item We construct a cost effective 3D MIMO channel sounding setup, where a virtual cylindrical array is created at the transmitter using a linear array and a mechanical rotor. The phase drift caused by increased measurement duration is corrected using a reference antenna.
\item We provide a technique to estimate the phase drift from the reference measurement, and correct for the phase errors in the channel measurements during post-processing. We also quantify the performance using simulations.
\item Using an iterative low complexity CLEAN algorithm, we extract the delay and directions of the MPCs, and validate them using the map of the environment. 
\item Using the UMa and UMi measurements, we characterize the elevation and azimuth spreads at the BS and UE for an O2I environment, and study their dependency on the BS and UE heights.
\item Using the measurements done with UEs on different floors, we study the feasibility of elevation beamforming in UMa and UMi environments.
\item Using the measured channel impulse responses, we examine the channel hardening aspects of Massive MIMO and the optimality of MRC combiners.
\eit 

The remainder of the paper is organized the following way: The architecture of the proposed channel sounder and the concept of the reference antenna is introduced in Sec.~\ref{sec:setup}. The measurement environment is described in Sec.~\ref{sec:envi}. Outlier suppression, the phase drift correction technique and the CLEAN algorithm for MPC parameter extraction are given in Secs.~\ref{sec:outlier},~\ref{sec:phase_correction}, and~\ref{sec:CLEAN}, respectively. The MPC parameters are validated for several measurement positions in Sec.~\ref{sec:mpc_validation}. Sec.~\ref{sec:res_spread} characterizes the azimuth and elevation spreads, Sec.~\ref{sec:ele_bf} provides some results on the feasibility of elevation beamforming and Sec.~\ref{sec:capacity} provides the capacity results. A summary and conclusions wrap up the paper in Sec.~\ref{sec:conclusions}. The mathematical details are relegated to the Appendix.  
\begin{figure} 
 \begin{center}
\includegraphics[width=0.8\linewidth]{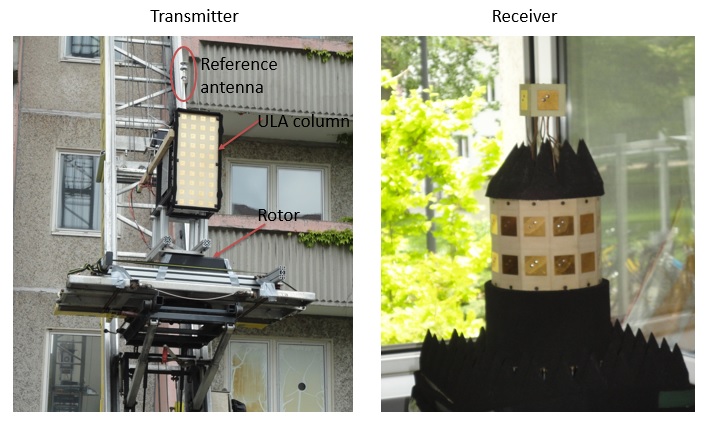}
\end{center}
\vspace{-1cm}
\caption{Transmit and Receive antenna array setup
\label{fig:Tx_Rx_setup}}
\end{figure}

\section{Measurement setup and environment}
\subsection{Measurement setup}\label{sec:setup}
The basic channel sounder setup is based on the switched-array principle (RUSK-HyEff sounder). A multi-tone signal generated at the transmitter (TX) is upconverted to the passband by a single RF chain, and connected by an electronic switch sequentially to the different elements of a physical antenna array. At the receiver (RX), similarly the different antenna elements are sequentially connected by an electronic switch to a single down-conversion chain, and the received signal is recorded for post processing, i.e., computation of the transfer function. This approach can provide the transfer function of all combinations of transmit and receive antenna elements while requiring only a single RF chain at TX and RX, and has thus been used extensively in the past~\cite{Thoma_2001,Thoma_2000,Sommerkorn_2014}.  

However, for the measurements in 3D, the number of TX antenna elements required for highly accurate measurements is considerably larger than the size of available electronic switches with reasonable price and attenuation. We used a hybrid setup that combined the switched array with a virtual array. Specifically, the transmit setup consists of a dual polarized 8 element (16 port) vertical uniform linear array (ULA), mounted on a programmable rotor. The transmit antenna array setup is shown in Fig.~\ref{fig:Tx_Rx_setup}. The actual array at the transmitter has more than 8 elements, where the excess elements are used as dummy to assuage any edge effects caused by mutual coupling. The antenna patch elements have 3~dB beamwidth of $100$~degrees in elevation and $26$~degrees in azimuth. To increase the gain of the ULA in azimuth, groups of 4 horizontal antenna elements formed a narrow transmit beam in azimuth ($\ie$, restricting azimuth opening angle) by using a pre-configured, controlled,  power divider (equal-split) array feeder network, constituting one ``effective'' element of the (vertical) array. 

The transmit RF chain is connected to the ULA using a 16 port electronic switch. The rotor is circularly scanned in the azimuth plane in steps of 6 degrees, thereby creating a virtual cylindrical antenna array structure comprising of 8 rings with 60 elements uniformly placed on each ring. This can be thought of as a vertically stacked polarimetric uniform cylindrical patch array (VSPUCPA). The advantage of this setup is that a 16 port switch is sufficient to measure the channel transfer function of a 480 element (960 port) cylindrical antenna structure. On the down side, the mechanical rotor introduces significant delay between the successive ULA column measurements which will be addressed later in the paper. 

At the receiver end, we use a physical cylindrical antenna array structure comprising of 2 rings with 12 dual polarized antenna elements uniformly placed on each ring, hence forming a stacked polarimetric uniform circular patch array (SPUCPA). The received signal is connected to the SPUCPA using a 48 port electronic switch. The receive antenna array setup is shown in Fig.~\ref{fig:Tx_Rx_setup}. Two rubidium (Rb) clocks, synchronized with respect to each other, were used at both transmitter and receiver ends. 

The transfer function measurements are done in the following sequence: For a given transmit antenna element, we complete the entire receive antenna array measurements before switching to next TX antenna element -- among the RX antenna elements, we loop over the 12 receiver positions on the ring and first take the measurement of upper ring elements followed by lower ring elements. We first complete the horizontally polarized port measurements for the receiver array followed by the vertically polarized port measurements. At the transmitter, we complete the measurements over the 8 element ULA column (vertically polarized port measurements followed by horizontally polarized port measurements) and the rotor moves to next position on the circle (60 rotor positions in total). Our measurement results directly give the channel transfer function uniformly sampled at $N_{\rm f}=257$ frequency points, uniformly sampled over the $2.52-2.54$~GHz band. The frequency band is chosen because of the available license in this band. To improve the signal-to-noise ratio (SNR) of the measurement, the measurements of all TX-RX antenna pairs was repeated 10 times at each rotor position. Each TX-RX antenna pair measurement takes $12.85$~$\mu$s. The switching delay at the transmitter and receiver is $12.85$~$\mu$s. Thus, it takes $1.23$~ms to complete a SIMO measurement between one TX element and the receiver array and $19.94$~ms to complete one snapshot of the measurement between a TX ULA column and the Rx antenna array at each rotor position. The rotor roughly takes $10$~s to move to the next position on the circle. Thus, the total MIMO measurement duration is roughly 10 minutes. The rotor movement and the subsequent start and stop of the MIMO measurements was controlled using a cellular (UMTS) connection. The measurement site and time was picked such that there were few (if any) moving scatterers in the environment.

Because of residual clock offset and the phase drift of the transmit and receive clocks, which is unavoidable even with the precision Rb clocks we used, the relative phase between the clocks drifts over time. The phase drift in the transfer function measurements, if not corrected, will introduce large errors in the direction estimates of the MPCs, which are critical for channel modeling. Since the phase drift errors accumulate over time, the errors introduced during switching delays are negligible relative to the error introduced during rotor delay. Using a reference measurement, we can measure the instantaneous phase drift for different rotor positions and use it to further correct the phase drift in the actual channel measurements. For this purpose, a vertically polarized omni-directional discone antenna was mounted in the center of structure of the VSPUCPA at a height of about $60$~cm above the chassis of the VSPUCPA, as indicated in Fig.~\ref{fig:Tx_Rx_setup}. It is connected to the $16^{\kth}$ port of the transmit switch (instead of the horizontally polarized port of the $8^{\kth}$ element of the ULA). The discone antenna is physically stationary, i.e., does not move with the rotor, and thus provides a reference signal that can be used for phase drift correction during post-processing. Similar concept of reference measurement for phase drift correction is also used in~\cite{Dunand_VTC_2007}, where SIMO measurements were conducted with a setup comprising of virtual planar array at the receiver end.

\subsection{Measurement environment}\label{sec:envi}
The measurements were carried out on the campus of Technical University Ilmenau, Germany, which has building structures typical of many European urban environments, with $3-5$ story buildings. The measurement map with TX and RX locations is shown in Fig.~\ref{fig:art_tx_mpcs}. The TX was placed on the roof of {\em House M}. The receiver was placed inside {\em House F}, which has Line-of-sight (LOS) to the transmitter. The RX was placed in different rooms on the second and fourth floor. The distance between TX and RX was $150$~m. Two sets of measurements were performed--one with TX placed 5~m above the rooftop of House M (macro-cellular) and one with TX placed 6~m below the rooftop (micro-cellular). A crane elevator was used in positioning the TX antenna array to each specific height, either for above rooftop or below rooftop. For micro-cellular measurements, the rotor movement was limited to a sector $-108$~deg to $108$~deg, with $0$~deg representing the transmitter ULA column roughly facing the house in which the receiver was placed. 

\section{Post Processing}\label{sec:post_proc}
\subsection{Outlier suppression and snapshot averaging}\label{sec:outlier}
For each TX-RX antenna pair, we take 10 channel transfer function measurements (snapshots). Under reasonable SNR conditions, all 10 snapshots should be perfectly correlated. However, it has been observed that for some of the TX-RX antenna pair measurements, a few snapshots had significantly different power compared to the rest. This might be because of the mechanical vibrations of the transmitter setup (the base supporting the rotor) or the occasional switching errors.  The latter explanation is more likely, as it has been observed that once in a while the switch gets stuck and there is a small delay in the switching. The false snapshots are usually very few in number, and can be eliminated using the outlier suppression technique given in Appendix~\ref{app:outlier}. The channel transfer functions are averaged only over the good snapshots. It was observed that the horizontal polarization measurements were relatively more affected by the switching errors, and hence only the vertical polarization measurements were used for further processing. All our post-processing and the results are based on single-polarimetric (vertical polarization) case.

For the further discussion, we define $H_r(f_k)$ as the $8\times 24$ matrix of channel transfer function between the 8-element transmit ULA column and the receiver array, measured at frequency $f_k$, when the rotor is at position $r (1\leq r\leq 60)$. Let $H_r^{\rm ref}(f_k)$ be the $1\times 24$ vector of channel transfer function between the transmit reference (discone) antenna and the receiver array, measured at frequency $f_k$, when the rotor is at position $r$.

\subsection{Phase drift}\label{sec:phase_correction}
As discussed earlier, because of the non-idealities of the clocks, the switching and rotor delays introduce phase drift errors in the MIMO channel transfer function measurements. The reference channel measurement can be used to estimate and correct for the phase drift errors due to rotor delay.

We now describe a phase drift estimation technique we ultimately used for the measurement evaluations. Other investigated phase drift estimation techniques, which performed worse, are summarized in Appendix~\ref{app:phase_drift_est_other}. 

\subsubsection{Phase drift estimation}
Since the switching delays are much smaller than the rotor delay, for now we ignore the phase drift errors due to switching delays and only correct for the phase errors due to rotor delays.

For each rotor position, $r$, we have a SIMO reference channel measurement between the reference antenna and the receiver array, $\ie$, $H_r^{\rm ref}(f_k)$. Since the transmit reference antenna has a uniform radiation pattern in the azimuth plane, and since the radiation pattern does not vary with the rotor position, the phase drift introduced by the rotor delay is reflected in the phase of the complex MPC path gains corresponding to respective SIMO reference channel. 

We use CLEAN to extract the parameters of the MPCs from the transfer function measurements. The algorithm is summarized in Sec.~\ref{sec:CLEAN}. Using~\eqn{eq:delay_angle_estimate}, the delay, azimuth of arrival (AoA), and elevation of arrival (EoA) of the strongest MPC, corresponding to the SIMO reference channel, when the rotor is in position $r$, is given by
\begin{equation}
\brac{\hat{\tau}^{(r)},\hat{\phi}_{\rm R}^{(r)},\hat{\psi}_{\rm R}^{(r)}}
=\!\arg\max_{\tau,\phi,\psi}\!\Bigg\arrowvert\!\!\sum_{k=1}^{N_{\rm f}} \exp\brac{-j2\pi f_k \tau} B_{\rm R}\brac{\phi,\psi}H_r^{\rm ref}(f_k)^{\dagger} \!\!\Bigg\arrowvert
\end{equation}
where $B_{\rm R}\brac{\phi,\psi}$ is the $1\times 24$ vector of calibrated vertically polarized beam pattern of the receiver array, for different azimuth and elevation angles; $(.)^{\dagger}$ denotes the hermitian transpose of the matrix and $|.|$ denotes the absolute value of a complex quantity. 

The estimated complex path gain of the strongest MPC, using~\eqn{eq:mpc_strength_est}, is given by
\begin{equation}\nonumber
\hat{\alpha}^{(r)}\!
=\!\brac{\!\!N_{\rm f}\! \abs{\abs{\!B_{\rm R}\brac{\hat{\phi}_{\rm R}^{(r)},\hat{\psi}_{\rm R}^{(r)}}\!}}^2}^{-1}
\sum_{k=1}^{N_{\rm f}}\! \exp\!\brac{j2\pi f_k \hat{\tau}^{(r)}}\!H_r^{\rm ref}(f_k) B_{\rm R}^{\dagger}\brac{\hat{\phi}_{\rm R}^{(r)},\hat{\psi}_{\rm R}^{(r)}}
\end{equation}
where $||.||$ denotes the Euclidean norm of the vector.

It has been observed that the delay and direction estimates obtained above were very similar (with up to $1$~degree variation) for all $60$ rotor positions and $\abs{\hat{\alpha}^{(r)}}$ roughly remained the same for different $r$, which indicates that there is no significant channel variations over the MIMO measurement duration. 

\subsubsection{Phase drift correction} 
For each rotor position, the corresponding reference channel measurement and the ULA column measurement experience similar
phase drift, since the drift is caused by the common clock. Thus, the phase drift estimated using the reference channel can be used to correct the phase in the actual channel transfer functions as given below  
\begin{align}
\tilde{H}_r(f_k)&\define H_r(f_k)\exp(-j\angle\hat{\alpha}^{(r)}), 1\leq k\leq N_{\rm f}, 1\leq r\leq 60\nonumber\\
\tilde{H}_r^{\rm ref}(f_k)&\define H_r^{\rm ref}(f_k)\exp(-j\angle\hat{\alpha}^{(r)}), 1\leq k\leq N_{\rm f}, 1\leq r\leq 60\nonumber
\end{align}
where $\angle (.)$ denotes the angle of the complex quantity.

Let $H(f_k)\define\sbrac{\tilde{H}_1(f_k)^T \tilde{H}_2(f_k)^T \cdots \tilde{H}_{60}(f_k)^T}^T$ be the phase corrected $N_{\rm T}(480)\times N_{\rm R}(24)$ matrix of MIMO channel transfer function, measured at frequency $f_k$. We apply CLEAN to these transfer functions and extract the delay and direction parameters of the MPCs. 

\subsubsection{Residual phase drift from switching delays}\label{sec:res_phase_err}
The above drift correction technique compensates for the phase drift due to rotor delay. We had implicitly assumed that the phase drift due to the switching delays do not impact the parameter estimates significantly. We will now quantify the error using simulations. For this, we generate the synthetic channel transfer functions using~\eqn{eq:chan_tranfun_model}~\footnote{Only a single scatterer is assumed in the environment. The scatter AoD and AoA are generated uniform from $[-180,180]$~deg, the EoD and EoA are generated uniform from $[-20,20]$~deg, the delay was generated uniform from $[150,300]$~m.}, the phase drift was modeled as a random-walk process and added to the transfer functions, as done in~\cite{Almers_2005}. For the simulations, we used the same setup as the actual measurement thereby incorporating the switching and rotor delays. The phase drift correction technique explained earlier, is applied and the MPC parameters are extracted from the phase corrected transfer functions. The impact of residual drift is quantified in terms of root mean square error (RMSE) in the directional estimates. For an Allan deviation (1s) of $10^{-10}$, which is typical of the clocks used in the measurement, the RMSE in azimuth of departure (AoD), elevation of departure (EoD), azimuth of arrival (AoA) and elevation of arrival (EoA) is $0.18$~deg, $0.10$~deg, $0.28$~deg and $2.34$~deg respectively. Please note that these values are relatively small compared to the typical angular spreads that we would expect.

\subsection{MPC Parameter extraction using 3D CLEAN}\label{sec:CLEAN}
CLEAN is an iterative deconvolution technique first introduced in~\cite{JAHogbom:1974} for the enhancement of the radio astronomical maps of the sky and widely used in microwave and UWB communities as an effective post-processing method for time-domain channel measurements. However, the principle can also be used to extract the delay and direction information from the channel transfer functions. It is a grid search based algorithm and hence the resolution is limited by the grid size. Angular grid size of 1 degree and delay grid size of $3$~m is used for all the MPC extractions in the paper. Although other high resolution algorithms such as RIMAX provide better resolution, they are relatively more sensitive to any model mismatches~\cite{Thoma_2004,Landmann_2012}.~\footnote{We noticed convergence issues with the RIMAX estimator when applied on the measurement data. This can be because of the mismatch in the data models, caused by the residual phase drift and switching errors. For this reason we used CLEAN for data processing. Since CLEAN is a single path estimator, it does not have convergence issues.} 

We now briefly describe the channel propagation model and the algorithm for parameter extraction: 
Let $B_{\rm T}\brac{\phi,\psi}$ and $B_{\rm R}\brac{\phi,\psi}$ respectively be the $N_{\rm T}\times 1$ and $1\times N_{\rm R}$ calibrated vertically polarized beam patterns of the transmit and receive arrays, for different azimuth and elevation angles. The $N_{\rm T}\times N_{\rm R}$ matrix of the channel transfer function between the transmit and receive arrays, measured at frequency $f_k$, can be modeled as sum of discrete MPCs as given below
\begin{equation}\label{eq:chan_tranfun_model}
H\brac{f_k}
=\!\sum_l \Big(\!\alpha_l B_{\rm T}\!\brac{\phi_{{\rm T},l},\psi_{{\rm T},l}}B_{\rm R}\!\brac{\phi_{{\rm R},l},\psi_{{\rm R},l}}
\exp\brac{-j2\pi f_k \tau_l}\Big)\!+\!N(f_k),1\leq k\leq N_{\rm f}
\end{equation}
where $\phi_{{\rm T},l}$ and $\psi_{{\rm T},l}$ respectively denote the Azimuth of departure (AoD) and Elevation of departure (EoD) for the $l^{\kth}$ MPC; $\phi_{{\rm R},l}$ and $\psi_{{\rm R},l}$ respectively denote the Azimuth of arrival (AoA) and Elevation of arrival (EoA); $\tau_l$ and $\alpha_l$ respectively denote the delay and complex path gain of the $l^{\kth}$ MPC; $N(f_k)$ is the receiver noise matrix with $\iid$ complex Gaussian entries. 

Assuming that the MPCs are resolvable,~\footnote{CLEAN is a single path estimator and cannot distinguish two MPCs with very similar delay and angles. Here we assume that two MPCs are separable in atleast one of the delay, transmit azimuth, transmit elevation, receive azimuth or receive elevation domains. If two MPCs have very similar parameters in all the domains, then estimating them as a single MPC is reasonable for angular and delay spread computations.} the joint maximum likelihood (ML) estimate of the delay and directions corresponding to the strongest MPC is given by
\begin{multline}\label{eq:delay_angle_estimate}
\brac{\hat{\tau}_1,\hat{\phi}_{{\rm T},1},\hat{\psi}_{{\rm T},1},\hat{\phi}_{{\rm R},1},\hat{\psi}_{{\rm R},1}}\\
=\arg\! \!\!\!\max_{\tau,\phi_{\rm T},\psi_{\rm T},\phi_{\rm R},\psi_{\rm R}}\!\Bigg\arrowvert\!\!\sum_{k=1}^{N_{\rm f}} \exp\brac{-j2\pi f_k \tau}
 B_{\rm R}\brac{\phi_{\rm R},\psi_{\rm R}}H(f_k)^{\dagger} B_{\rm T}\brac{\phi_{\rm T},\psi_{\rm T}}\Bigg\arrowvert
\end{multline}

The estimated path gain of the corresponding MPC is given by
\begin{multline}\label{eq:mpc_strength_est}
\hat{\alpha}_1\!
=\!\brac{\!\!N_{\rm f}\! \abs{\abs{\!B_{\rm T}\brac{\hat{\phi}_{{\rm T},1},\hat{\psi}_{{\rm T},1}}\!}}^2\!\abs{\abs{\!B_{\rm R}\brac{\hat{\phi}_{{\rm R},1},\hat{\psi}_{{\rm R},1}}\!}}^2}^{-1}\\
\times\sum_{k=1}^{N_{\rm f}} \exp\brac{j2\pi f_k \hat{\tau}_1}B_{\rm T}^{\dagger}\brac{\hat{\phi}_{{\rm T},1},\hat{\psi}_{{\rm T},1}}H(f_k) B_{\rm R}^{\dagger}\brac{\hat{\phi}_{{\rm R},1},\hat{\psi}_{{\rm R},1}}
\end{multline}

The contribution of thus estimated strongest MPC is subtracted from the transfer functions.
\begin{equation}\label{eq:multipath_intf_cancel}
\tilde{H}(f_k) = H(f_k) -\bigg(\hat{\alpha}_1\exp\brac{-j2\pi f_k\hat{\tau}_1}
B_{\rm T}\!\brac{\hat{\phi}_{{\rm T},1},\hat{\psi}_{{\rm T},1}}B_{\rm R}\!\brac{\hat{\phi}_{{\rm R},1},\hat{\psi}_{{\rm R},1}}\bigg)
\end{equation}
The parameters of the next strongest MPC are extracted from the new transfer functions, using~\eqn{eq:delay_angle_estimate} and~\eqn{eq:mpc_strength_est}, and the corresponding contribution is subtracted using~\eqn{eq:multipath_intf_cancel}. This process repeats until the absolute path gain of thus extracted MPC falls below a pre-determined threshold. 

\subsubsection{Iterative evaluation of~\eqn{eq:delay_angle_estimate}} It can be seen that~\eqn{eq:delay_angle_estimate} requires a five dimensional grid search which is computationally intense. However, we reduced the effort by using an iterative three dimensional grid search as described below:~\footnote{This approach is similar to the SAGE implementation~\cite{Fleury_sage} which uses an iterative one dimensional grid search technique. To improve the accuracy of parameter estimates, we instead use joint three dimensional grid search.}

Let $\hat{\phi}_{\rm T}^{(i)}$  and $\hat{\psi}_{\rm T}^{(i)}$ be the current transmit angular estimates. The joint estimate of the delay and the receive angular estimates now simplifies to
\begin{equation}
\brac{\hat{\tau},\hat{\phi}_{\rm R}^{(i+1)},\hat{\psi}_{\rm R}^{(i+1)}}
=\arg\max_{\tau,\phi,\psi}\Bigg\arrowvert\sum_{k=1}^{N_{\rm f}}\! \exp\brac{-j2\pi f_k \tau}
B_{\rm R}\brac{\phi,\psi}H(f_k)^{\dagger} B_{\rm T}\brac{\hat{\phi}_{\rm T}^{(i)},\hat{\psi}_{\rm T}^{(i)}}\Bigg\arrowvert
\end{equation}
Using this receive angular estimates, we refine the joint estimate of delay and the transmit angular estimates in the next iteration, as given below
\begin{equation}
\brac{\hat{\tau},\hat{\phi}_{\rm T}^{(i+2)},\hat{\psi}_{\rm T}^{(i+2)}}
=\arg\max_{\tau,\phi,\psi}\Bigg\arrowvert\sum_{k=1}^{N_{\rm f}}\! \exp\brac{-j2\pi f_k \tau}
 B_{\rm R}\brac{\hat{\phi}_{\rm R}^{(i+1)},\hat{\psi}_{\rm R}^{(i+1)}}H(f_k)^{\dagger} B_{\rm T}\brac{\phi,\psi}\Bigg\arrowvert
\end{equation}
From the simulations, we observed that 3 iterations are sufficient for the convergence of the parameters.

{\em Initialization:} The initial transmit angular estimates are obtained by doing beam-forming at the transmitter.~\footnote{In general, initialization can be done by either transmit or receive angular estimates. Since $N_{\rm T}>>N_{\rm R}$ for our measurement setup, we initialize the algorithm by transmit angular estimates.} The AoD and EoD corresponding to the strongest MPC as seen by the transmit antenna array structure is
\begin{equation}
\brac{\hat{\tau},\hat{\phi}_{\rm T}^{(0)},\hat{\psi}_{\rm T}^{(0)}}=\arg\max_{\tau,\phi,\psi}\max_n\Bigg\arrowvert\sum_{k=1}^{N_{\rm f}} \exp\brac{-j2\pi f_k \tau}
 H^{(n)}(f_k)^{\dagger} B_{\rm T}\brac{\phi,\psi}\Bigg\arrowvert
\end{equation}
where $H^{(n)}(f_k)$, the $n^{\kth}$ column of matrix $H(f_k)$, is the channel transfer function vector between the transmitter array and the $n^{\kth}$ receive antenna.

\subsubsection{Dynamic range limitation} Since CLEAN is a grid search based technique, the estimated MPC parameters can be different from the true parameters up to the nearest grid point. Hence the slight mismatch in parameter estimates can result in imperfect cancellation of MPC contributions. The residual will appear as a ghost MPC, whose delay and direction parameters are similar to original MPC and the path strength of the ghost MPC is typically much smaller than the original MPC. While the ghost MPCs will not affect the azimuth angular spreads, they can distort the elevation spreads at the transmitter as the typical elevation spreads are much smaller than the azimuth spreads. Using the synthetic data, we observed that the ghost MPC power is typically $20$~dB below the original MPC (for the beam pattern and grid size used in our measurement setup). Hence, for all our evaluations, we only extract the MPCs within $20$~dB of the strongest MPC in order to minimize the impact of ghost MPCs.  

\section{MPC extraction results and validation}\label{sec:mpc_validation}
We now present the parameter extraction results, and validate them, for a sample UMa and UMi measurements. 
\subsection{Macrocell measurement}\label{sec:mpc_validation_art}
The transmitter was placed 5~m above the rooftop of house M and the receiver was placed inside the room 2300 of house F, which  is on the second floor. The receiver location can be seen in Fig.~\ref{fig:brt_rx_mpcs}. The room door was closed during the measurement. The transfer function measurements were corrected for the phase drift and the MPC parameters were extracted using CLEAN. 

\begin{table}\center
  \caption{MPC parameters for a macrocell measurement}
  \begin{tabular}{|l|l|l|l|l|l|p{4cm}|}
    \hline
  MPC & EoD & AoD & EoA & AoA & Power(dB) & Description \\ \hline
     1  &   6  &   1  & -10  &   2 & -142.50 & LOS\\ \hline
     2  &   6  &  22  & -11  &   1 & -151.73 & Rooftop diffraction at House B\\ \hline
     3  &   4  &   1  &  -6  &  30 & -154.26 & Side wall reflection at Rx\\ \hline
     4  &   6  &   2  &  -2  & -49 & -156.53 & Side wall reflection at Rx\\ \hline
     5  &   5  & -15  & -12  &  -2 & -156.65 & Reflection from House C\\ \hline
     6  &   6  &  45  & -11  &  -2 & -156.65 & MPC from Setup\\ \hline
     7  &   9  & -22  &  -4  &  17 & -157.01 & Reflection from House D\\ \hline
     8  &   7  &   0  &  15  & 132 & -157.95 & Back wall refection at Rx\\ \hline
     9  &   4  &-147  &   1  &  18 & -158.76 & Refection from storage room\\ \hline
    10  &   5  &  28  & -10  &   3 & -159.02 & Rooftop diffraction at House A\\ \hline
    11  &   6  & -42  & -11  &   0 & -159.14 & MPC from setup\\ \hline
    12  &   5  &  10  &   0  &  16 & -159.48 & Reflection from House B at Tx\\ \hline
    13  &   5  &   3  & -17  & 172 & -159.58 & Back wall refection at Rx\\ \hline
    14  &   7  &   3  &   2  &  14 & -160.19 & Diffraction from window at Rx\\ \hline
    15  &  10  &   4  &  14  &  37 & -161.88 & Sidewall reflection at Rx\\ \hline
    16  &   7  &   0  &  14  &  82 & -162.03 & Sidewall reflection at Rx\\ \hline
\end{tabular}
\label{tab:art_mpcs}
\end{table}
The MPC parameters are tabulated in Table~\ref{tab:art_mpcs}.~\footnote{Following 3GPP tradition, we consider the downlink, so that BS becomes synonymous with transmitter/departure, and UE with receiver/arrival.} All angles are given in degrees. The MPCs are sorted in terms of the path power. Sign convention: A positive elevation at the TX/RX represents ray pointing to the ground and a negative elevation represents ray pointing towards the sky. The MPC parameters are validated using the map of the propagation environment. {From the distances obtained using Google earth, we compute the relative azimuth and elevation angles under which the TX and RX sees the possible scatterers (buildings, rooftops etc.) in the environment and the corresponding overall path lengths associated with that scattering. The extracted MPCs are then matched with the physical scatterers based on the delay, EoD, AoD, EoA, and AoA.} The possible interaction processes with the environment are listed for each MPC. 

\begin{figure}
 \begin{center}
\includegraphics[width=0.8\linewidth]{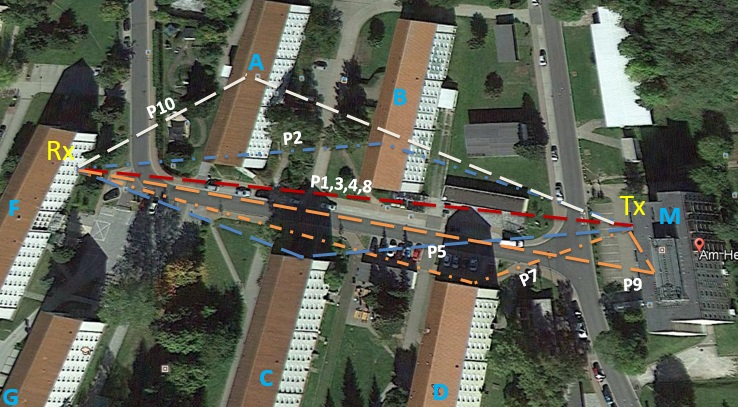}
\end{center}
\vspace{-1cm}
\caption{Figure showing the direction of departure of MPCs for a Macrocell measurement. 
\label{fig:art_tx_mpcs}}
\end{figure}

Sample MPCs at the TX side are shown in Fig.~\ref{fig:art_tx_mpcs}. The MPCs at the RX side can be found in the conference paper~\cite{kristem_globecomm}. To avoid cluttering in the figure, MPCs with similar direction of arrivals/departures are shown together. For instance P1, P3, P4, and P8 have similar direction of departure, but they have different direction of arrival. They are shown using different line styles -- P1 correspond to LOS, P8 correspond to back wall reflection, P3 and P4 correspond to side wall reflections. Similarly MPCs P7 and P9 have similar direction of arrival, but different direction of departure -- P7 correspond to reflection from house D and P9 correspond to reflection from storage room. While the table lists all the MPCs within $20$~dB dynamic range, only the strongest few MPCs are shown in the map. Apart from the LOS, the dominant propagation at the transmitter were the reflections from the houses and the rooftop diffraction. There were reflections from the storage room located to the back of the transmitter on house M (for instance P9). The dominant propagation at the receiver was the reflections from back and side walls and the diffraction at the window corners. There were a few MPCs that were not consistent with the map of the environment (P6 and P11). These MPCs consistently showed up for all measurement points and were identified as being caused by the setup -- they were scattering from the corners of the square base on which the TX array and the rotor were mounted.

\begin{figure} 
 \begin{center}
\includegraphics[width=0.8\linewidth]{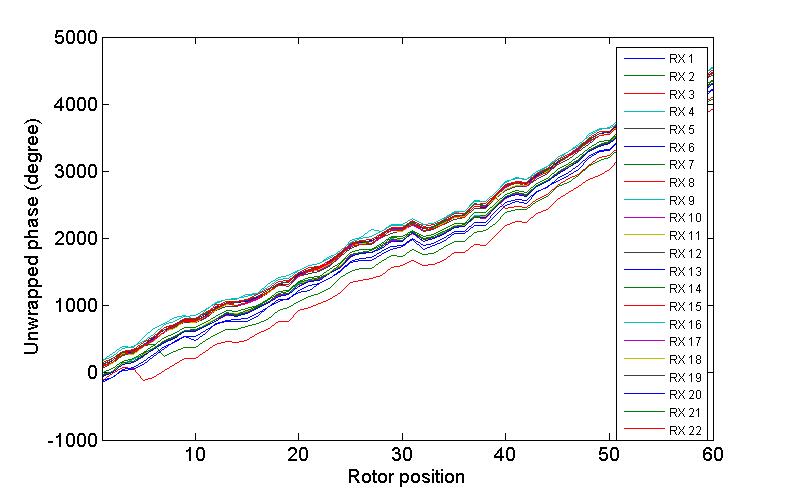}
\end{center}
\vspace{-1cm}
\caption{Phase variation in the reference channel before drift correction, for a macrocell measurement.
\label{fig:ART_meas_phase_bef}}
\end{figure}

\begin{figure} 
 \begin{center}
\includegraphics[width=0.8\linewidth]{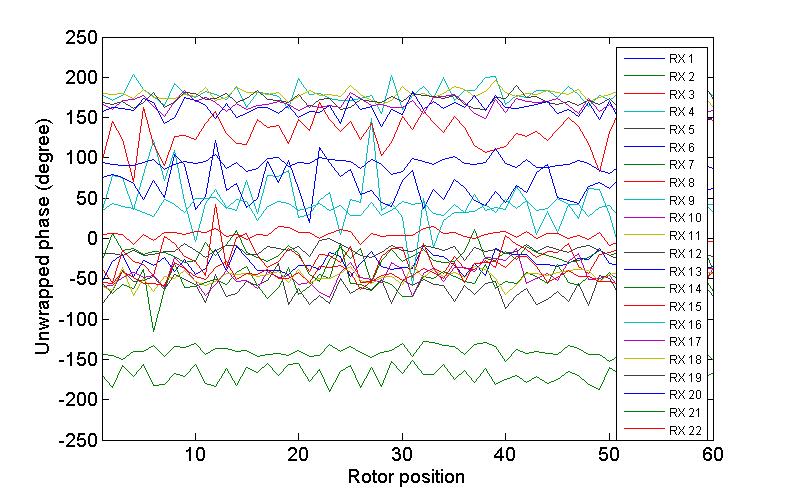}
\end{center}
\vspace{-1cm}
\caption{Phase variation in the reference channel after drift correction, for a macrocell measurement.
\label{fig:ART_meas_phase_after}}
\end{figure}

We now demonstrate the effectiveness of the phase drift correction technique described in Sec.~\ref{sec:phase_correction} by plotting the phase variation in the reference channel measurements before and after phase drift correction. Fig.~\ref{fig:ART_meas_phase_bef} plots the phase in the reference channel transfer functions, $\ie$, $\angle(\sum_{k=1}^{N_{\rm f}}H_r^{\rm ref}(f_k))$. The phase variation with rotor position $r$ is shown for $N_{\rm R} (24)$ receive antennas. If there were no phase drift, the phase in the reference channel would have remained the same for all rotor positions. Also, it can be seen that all the Rx antennas exhibit similar phase variation. This is because the rotor delay is significantly larger than the switching delays and mainly contribute to the phase drift. Fig.~\ref{fig:ART_meas_phase_after} plots the phase in the phase drift corrected reference channel transfer functions, $\ie$, $\angle(\sum_{k=1}^{N_{\rm f}}\tilde{H}_r^{\rm ref}(f_k))$. Since we can only correct for the phase drift due to rotor delays, there is a residual phase variation over the rotor positions. The RX antenna patches with higher signal-to-noise ratio (SNR) have smaller residual phase variations.~\footnote{Please note that we plot the measured phase (not the phase drift) and hence the phase  offset between different RX antennas correspond to the phase of the RX array beam patterns along the direction of strongest MPC.}

We use synthetic data to further validate the correctness of the MPCs obtained above and to quantify the impact of the measured residual phase drift on the parameter estimates. For this, we generate the synthetic channel transfer functions using~\eqn{eq:chan_tranfun_model}. For simplicity, let $H_{\rm syn}^{r,n_{\rm R}}(f_k)$  be the synthetic transfer function between TX ULA column corresponding to rotor position $r$ and the RX antenna element $n_{\rm R}$. Let $\theta_{n_{\rm R}}(r)$ be the residual phase drift in the reference channel, corresponding to rotor position $r$ and RX antenna element $n_{\rm R}$. 
The residual phase drift is added to synthetic transfer functions as given below
\begin{equation}\nonumber
H_{\rm syn}^{r,n_{\rm R}}(f)
=\!H_{\rm syn}^{r,n_{\rm R}}\!(f)\exp\!\brac{j\theta_{n_{\rm R}}(r)\!-\!j\theta_{n_{\rm R}}(1)}, 1\!\leq\! r\!\leq\! 60, 1\!\leq\!n_{\rm R}\!\leq\! N_{\rm R}
\end{equation}

The MPC parameters are extracted using CLEAN. An RMSE of $1.28$~deg, $0.11$~deg, $0.59$~deg and $2.29$~deg was observed for AoD, EoD, AoA and EoA respectively. These numbers are comparable to the numbers obtained in Sec.~\ref{sec:res_phase_err}.~\footnote{Please note that in Sec.~\ref{sec:res_phase_err} we model the phase drift as a random walk process and quantify the impact of switching and rotor delays. Here, we use the measured residual phase drift (although observed in the reference channel) to quantify the error in the parameter estimates.} 

We also quantify the error in the angular spreads, which is a key parameter of interest. For this, we use the extracted MPCs (Table.~\ref{tab:art_mpcs}) as the true channel parameters and generate the synthetic channel transfer functions; the residual phase drift (measured using reference channel) is added to the transfer functions as done earlier, and the MPC parameters are extracted. The percentage error in the angular spreads, computed using true channel parameters and the estimated parameters, is $5.2\%$, $0.8\%$, $10.8\%$ and $12.1\%$ respectively for transmit azimuth, transmit elevation, receive azimuth and receive elevation. 

\subsection{Microcell Measurement}\label{sec:mpc_validation_brt}
\begin{figure}
 \begin{center}
\includegraphics[width=0.8\linewidth]{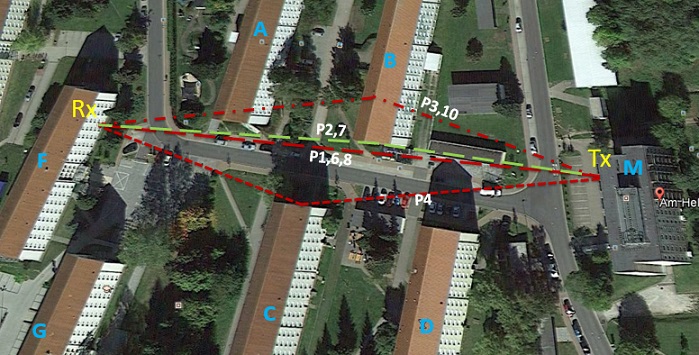}
\end{center}
\vspace{-1cm}
\caption{Figure showing the direction of departure of MPCs for a Microcell measurement. 
\label{fig:brt_tx_mpcs}}
\end{figure}

\begin{figure}
 \begin{center}
\includegraphics[width=0.8\linewidth]{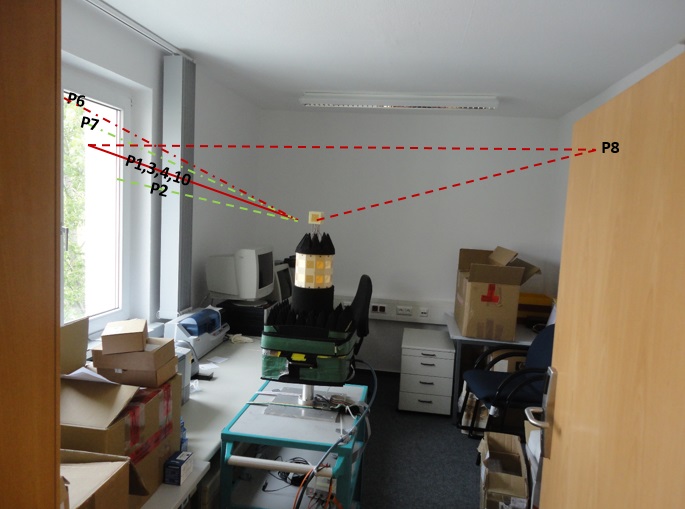}
\end{center}
\vspace{-1cm}
\caption{Figure showing the direction of arrival of MPCs for a Microcell measurement. 
\label{fig:brt_rx_mpcs}}
\end{figure}

The transmitter was placed 6~m below the rooftop of house M and the receiver was placed inside the room 2300 of house F. Sample MPCs at the TX and RX side are shown in Fig.~\ref{fig:brt_tx_mpcs} and~\ref{fig:brt_rx_mpcs} respectively~\footnote{The tabulated MPCs can be found in the conference paper~\cite{kristem_globecomm}.}. Once again the LOS, reflections from the near by houses and the roof top diffraction are the dominant propagation mechanisms. Because we limited the transmission directions to a sector, no reflections were observed from the back of the transmitter. 

\begin{figure} 
 \begin{center}
\includegraphics[width=0.8\linewidth]{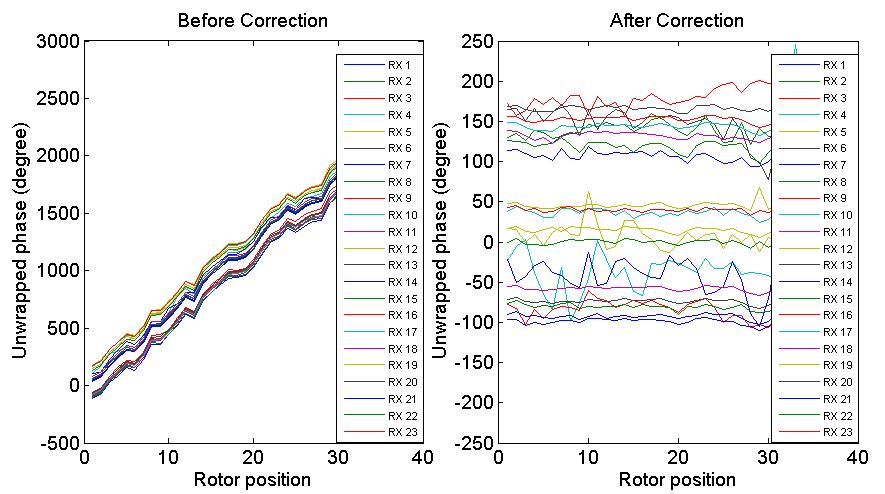}
\end{center}
\vspace{-1cm}
\caption{Phase variation in the reference channel before and after drift correction, for a microcell measurement.
\label{fig:BRT_meas_phase_bef_after}}
\end{figure}

Fig.~\ref{fig:BRT_meas_phase_bef_after} plot the phase variation in the reference channel transfer functions, before and after phase drift correction. The residual phase drift when tested using the synthetic data gave an RMSE of $1.89$~deg, $0.21$~deg, $0.88$~deg and $2.50$~deg respectively for AoD, EoD, AoA and EoA respectively. The resulting errors in the angular spreads is $6.4\%$, $8.5\%$, $2.5\%$ and $11.4\%$ respectively for transmit azimuth, transmit elevation, receive azimuth and receive elevation. 

\section{Results}
\subsection{Angular Spread Statistics}\label{sec:res_spread}
We now provide the angular spread statistics for the measurements done on different floors, for macrocell and microcell environments. For each measurement point, the root mean square (RMS) angular spread is computed from the extracted MPCs. We discard the MPCs scattered by the setup itself, as they do not represent the measurement environment. For macrocell, we restrict our evaluations to a sector of $-120$~deg to $120$~deg at the transmitter.~\footnote{We recompute the MPC parameters taking only the channel transfer function measurements done for the ULA columns corresponding to rotor position from $-120$~deg to $120$~deg. {Equivalently, we consider only the channel transfer functions measured with 41 out of the 60 ULA columns. We discard the channel transfer functions measured for the ULA columns facing the storage room, as those do not correspond to a realistic deployment scenario.}} This is because of the reflections observed from the storage room located at the back of transmitter (as observed in Sec.~\ref{sec:mpc_validation_art}), which heavily skews the azimuth spreads at the transmitter. In a typical macrocell setup, we do not expect reflections from the back of the base station. 

Furthermore, because of the switching errors and the residual phase drift in the measurement data, we only consider the subset of measurement points for which (i) the MPC parameters are in agreement with the map of the environment and (ii) the measured residual phase drift (in the reference channel), when tested on the synthetic data, gives RMSE of less than 2.5 degree in the AoD, and less than 0.5 degree in the EoD. After the filtering, we had 5 measurement points each for the second and fourth floors in the macrocell setup, and 6 and 4 measurement points for the second and fourth floors respectively in the microcell setup.

Table~\ref{tab:spread_art} lists the mean values of azimuth spread of departure (ASD), elevation spread of departure (ESD), azimuth spread of arrival (ASA) and elevation spread of arrival (ESA) for the macrocell and microcell environments respectively. The averaging is done over the measurements performed on the same floor. As expected, the elevation spreads are significantly smaller than the azimuth spreads both at the BS and UE. The elevation and azimuth spreads at the BS are smaller than the corresponding angular spread at the UE. This is because of more uniform scattering in the indoor environment. The measured ESD values are comparable to ESD of $0.79$--$1$ degree reported in~\cite{Rui_2014} (obtained using ray tracing simulations) for different UE heights, when BS-UE ground separation is $150$~m. The measured ESA values are smaller than the ESA of $5$~degree observed in~\cite{Rui_2014}. 
{The small ESA values occur mainly because the majority of MPCs propagate to the UE through a glass window with small elevation opening, as indicated in Fig.~\ref{fig:brt_rx_mpcs}, which is not typically modeled in ray tracing.}

{\em Angular spread dependency on BS/UE height: }It can be seen that the elevation spreads at the BS and UE decreases as the BS-UE height difference reduces: $\overline{\text{ESD}}$ and $\overline{\text{ESA}}$ decrease when the BS height is reduced (macrocell to microcell) for a given UE location, and they decrease when the UE is moved from second to fourth floor for a fixed BS location.  This is because of the increased power in the LOS component, when the BS and UE are aligned. While the power in the non-LOS MPCs also increase, not as much as the LOS component, there by resulting in the smaller elevation spreads. Similar observations of elevation spread variation with BS and UE height were also reported in~\cite{Rui_2014,Kalliola_2002,Zhang_2015_WCNC}. No clear patterns are observed for the azimuth spreads at the BS and UE. 

\begin{figure}
 \begin{center}
\includegraphics[width=0.78\linewidth]{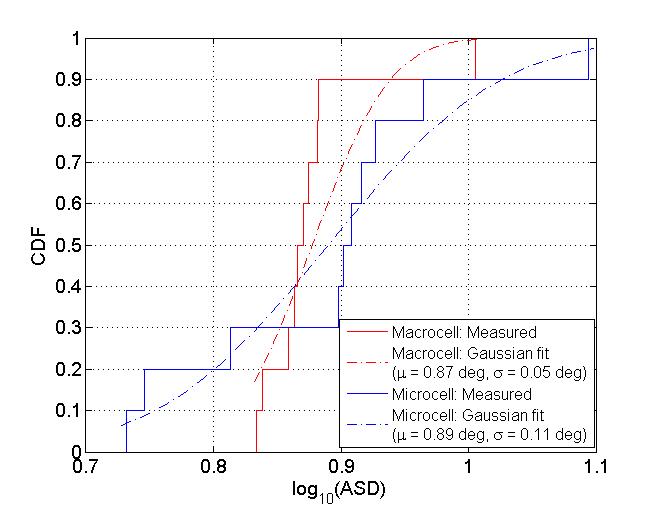}
\end{center}
\vspace{-1cm}
\caption{CDF of the BS azimuth spread for macrocell and microcell.
\label{fig:azimuth_spread}}
\end{figure}

\begin{figure}
 \begin{center}
\includegraphics[width=0.78\linewidth]{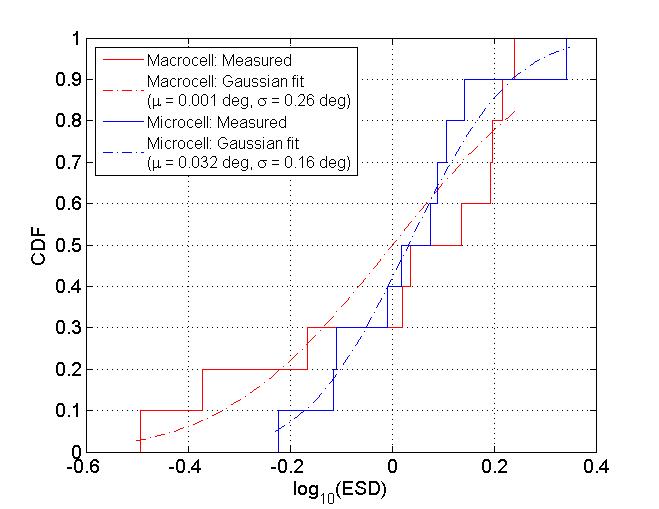}
\end{center}
\vspace{-1cm}
\caption{CDF of the BS elevation spread for macrocell and microcell.
\label{fig:elevation_spread}}
\end{figure}

{\em Distribution of BS angular spreads: }Fig.~\ref{fig:azimuth_spread} and~\ref{fig:elevation_spread} respectively plot the CDF of the elevation and azimuth spreads, combined over the  measurements done with UE on different floors. We see that the data fits reasonably well with the log-normal distribution~\footnote{The goodness of the fit is verified by applying KS test at 5\% significance level. }, and is consistent with the observations in~\cite{Rui_2014,Zhang_2015_WCNC, Zhang_2015_ICNC}. However, the observed angular spreads are smaller than the ones reported in these works. The massive antenna array structure used at the BS in our setup helps resolving the closely spaced MPCs in the azimuth and elevation, and hence resulting in smaller angular spreads at the BS.

\begin{figure}
 \begin{center}
\includegraphics[width=0.8\linewidth]{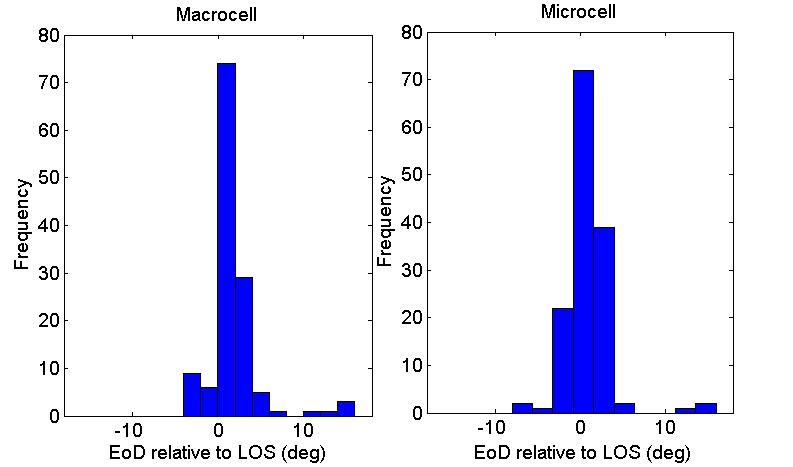}
\end{center}
\vspace{-1cm}
\caption{BS elevation angles, relative to LOS, for macrocell and microcell.
\label{fig:elevation_angles}}
\end{figure}

{\em Distribution of BS elevation angles: }Fig.~\ref{fig:elevation_angles} plots the histogram of the EoD of the MPCs (relative to the geometric LOS) for the macrocell and microcell scenarios. It can be seen that the distribution is skewed right, and more so for the microcell scenario. Since the major propagation scenarios are the rooftop diffractions and reflections along the building, most of the MPCs have larger elevation angles compared to LOS, thereby resulting in skewed distribution. Our observations are slightly different from the earlier works in the literature~\cite{Zhang_2015_WCNC, Zhang_2015_ICNC,Kavya_MILCOM_2013}, where the EoA was fitted using laplacian distribution.

\begin{table}\center
  \caption{Mean angular spreads (in degrees) at the BS and UE}
  \begin{tabular}{|l|p{2.6cm}|l|l|l|l|}
    \hline
BS Environment & UE location (Floor) & $\overline{\text{ESD}}$ & $\overline{\text{ASD}}$ & $\overline{\text{ESA}}$ & $\overline{\text{ASA}}$ \\ \hline
\multirow{2}{*}{Macrocell} 
 &    Second  &   1.24  &   7.79  &  3.34  &  22.77 \\ 
 &    Four    &   1.04  &   7.38  &  2.55  &  24.41 \\ \hline
\multirow{2}{*}{Microcell} 
 &    Second  &   1.15  &   7.62  &  2.53  &  20.57 \\  
 &    Four    &   0.78  &   8.54  &  2.22  &  33.88 \\ \hline
    \end{tabular}
\label{tab:spread_art}
\end{table}
\subsection{Elevation Beamforming}\label{sec:ele_bf}
One of the key advantages of 3D MIMO is the capability of separating two or more users by forming beams in the elevation domain at the transmitter. Two users can be separated in the elevation domain, if there is a reasonable non-overlapping region in the respective power density spectrum. While the power density spectrum also depends on the beam patterns of the transmit antenna array, the elevation spectrum of the channel alone is a good indicator. Ref.~\cite{Sommerkorn_2014} provides the following measure for the separability of users $i$ and $j$, based on the channel elevation spectrum of the users. 
\begin{equation}
r_{ij}\define\abs{\overline{\text{EoD}}_i-\overline{\text{EoD}}_j}-\text{ESD}_i-\text{ESD}_j
\end{equation}
where $\overline{\text{EoD}}$ is the power weighted mean EoD of the MPCs and ESD is the corresponding measurement's RMS elevation spread at the transmitter. The larger the $r_{ij}$, the smaller is the overlap in the elevation spectrum of the users $i$ and $j$, and the better the separability. In~\cite{Sommerkorn_2014}, $r_{ij}=0$ is used as a threshold for the separability of users.

For our macrocell measurements, we had 25 realizations of $r_{ij}$, with one UE placed on second floor and the other on the fourth floor. $r_{ij}$ was positive for 11 ($44\%$) realizations. Similarly, for the microcell measurements, $r_{ij}$ was positive for 13 out of 24 ($54\%$) realizations. Because of the smaller elevation spreads for the microcell measurements, the overlap in the elevation power spectrum of the UEs is small, and hence higher are the odds of separating users in the elevation domain.

\subsection{Uplink Capacity Analysis for Massive MIMO}\label{sec:capacity}
In theory, it is well known that as the number of antennas at the BS increases, $\ie$, in the massive MIMO regime,  the channel becomes (nearly) deterministic and the effect of small-scale fading is averaged out~\cite{marzetta_2010}. This has two fold implications: (i) For a BS serving single user, the channel capacity does not vary with different small-scale fading realizations and (ii) For a BS serving multiple users, the Maximum Ratio Combining (MRC) is as good as Zero-Forcing (ZF) and Minimum Mean Square Error (MMSE) combiner. In this section, we examine these two aspects with the measured channel impulse responses, for different antenna array structures.

{\em Capacity with different antenna array structures at BS:} We first examine the uplink channel capacity for a case of BS serving a single UE, for different antenna array structures at the BS. Here, the UE is assumed to be equipped with a single isotropic transmit antenna. 
{ The $N_{\rm T}\times 1$ channel frequency responses between the transmit antenna at the UE and the receive antenna array at the BS are constructed using the MPCs extracted from the measurement, as 
\begin{equation}\label{eq:chan_tranfun_gen}
H\brac{f_k}
=\!\sum_{l=1}^L \!\abs{\alpha_l} \exp\brac{j\theta_l}B_{\rm T}\!\brac{\phi_{{\rm T},l},\psi_{{\rm T},l}}
\exp\brac{-j2\pi f_k \tau_l},1\leq k\leq N_{\rm f}
\end{equation}

where the parameters $\cbrac{\alpha_l,\tau_l,\phi_{{\rm T},l},\psi_{{\rm T},l}}_{l=1}^L$ are the magnitude, delay and directions of the MPCs extracted from the measured channel transfer functions. $\cbrac{\theta_l, 1\leq l\leq L}$ are the phase of the MPCs which are generated $\iid$ uniform distributed in $\sbrac{0,2\pi}$. $B_{\rm T}(.,.)$ is the beam-pattern of the antenna array at the BS. Assuming an OFDM system, the uplink channel capacity is given by
\begin{equation}
C=\frac{1}{N_{\rm f}}\sum_{k=1}^{N_{\rm f}}\log_2\brac{1+\frac{1}{N_{\rm T}}\frac{H\brac{f_k}^{\dagger}H\brac{f_k}}{N_0}}
\end{equation}
where the noise power, $N_0$, is computed from the noise-only region of the channel impulse responses.

We consider two different antenna array structures at the BS: (i) 480-element cylindrical antenna array, as used in the channel measurements and (ii) $N_1\times N_2$ rectangular antenna array ($N_1$ elements in the azimuth plane and $N_2$ elements in elevation plane), with $\frac{\lambda}{2}$ spacing between the antenna elements. By varying the phase of the extracted MPCs (phase is generated $\iid$ across MPCs and realizations), we generate 300 small scale fading realizations of $H$, and hence the capacity. 
}

\begin{figure}
 \begin{center}
\includegraphics[width=0.78\linewidth]{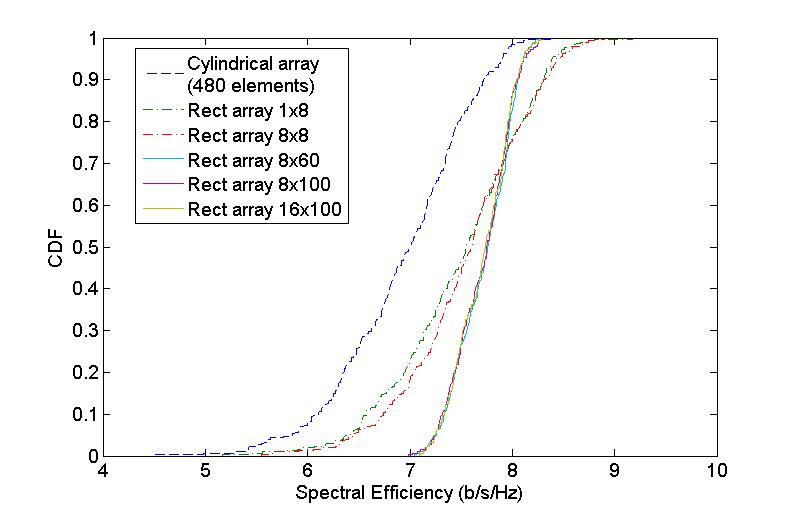}
\end{center}
\vspace{-1cm}
\caption{CDF of the uplink channel capacity for a sample macrocell measurement, for a BS serving single UE.
\label{fig:cap_cdf_single_ue}}
\end{figure}

Fig.~\ref{fig:cap_cdf_single_ue} plots the empirical CDF of the channel capacity, computed using the MPCs extracted for a sample macrocell measurement. It can be seen that the mean capacity with a cylindrical array is small compared to a rectangular array structure. Since the antennas used in the measurements have a narrow beamwidth in the azimuth, the effective number of antennas facing the receiver is small for cylindrical structures and hence a lower SNR gain relative to the rectangular array structures. For the rectangular array, as the number of antenna elements increased, the capacity CDF becomes steep because of the increased diversity order. Beyond rectangular array of 8x60, the capacity CDF did not change with the increase in the number of antennas. Even with 16 elements in the elevation and 100 elements in the azimuth, the capacity varied from 7.1 b/s/Hz to 8.2 b/s/Hz, over the realizations of small scale fading. 

\begin{figure}
 \begin{center}
\includegraphics[width=0.78\linewidth]{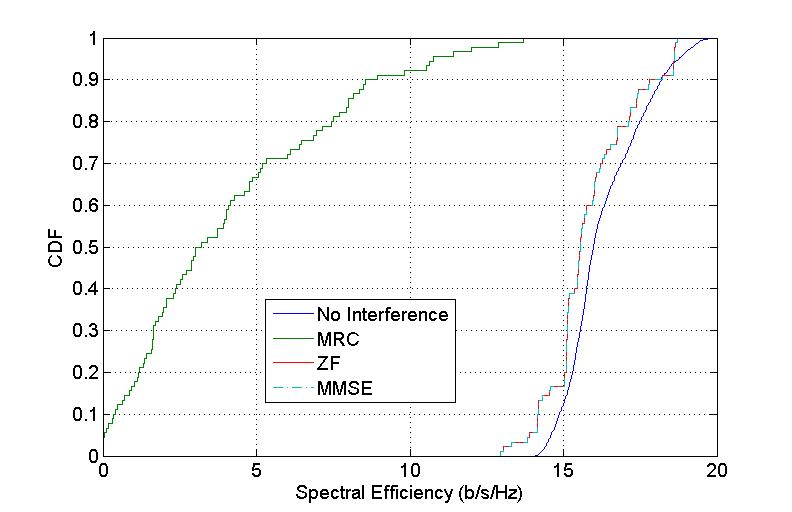}
\end{center}
\vspace{-1cm}
\caption{CDF of the uplink channel capacity for a macrocell scenario, for a BS serving two UEs.
\label{fig:cap_cdf_intf_case}}
\end{figure}

{\em Capacity with different receivers in interference scenario:} We now examine the uplink channel capacity for a case when a BS is serving two UEs located in different rooms. 
{ For instance, let $UE_i$ be the serving UE and $UE_j$ be the interference UE.  We assume that the BS has perfect knowledge of the channel. The uplink channel capacity for a linear combining receiver is given by
\begin{equation}
C=\frac{1}{N_{\rm f}}\sum_{k=1}^{N_{\rm f}}\log_2\brac{1+\frac{1}{N_{\rm T}}\frac{\abs{H_i\brac{f_k}^{\dagger}W_k}^2}{\abs{H_j\brac{f_k}^{\dagger}W_k}^2+\abs{\abs{W_k}}^2 N_0}}
\end{equation}
where $H_i\brac{f_k}$ is the $N_{\rm T}\times 1$ channel transfer function between the BS and the $UE_i$, generated using~\eqn{eq:chan_tranfun_gen}, based on the MPCs extracted from the measurements.  

The $N_{\rm T}\times 1$ weight vector for different linear detectors is given by~\cite{Molisch2010WirelessBook,marzetta_2010}
\[W_k =
  \begin{cases}
    H_i\brac{f_k}       & \quad \text{for MRC }\\
    \sbrac{G_k\brac{G_k^{\dagger}G_k}^{-1}}_1  & \quad \text{for ZF } \\
    \sbrac{G_k\brac{G_k^{\dagger}G_k+\frac{1}{N_0}I}^{-1}}_1  & \quad \text{for MMSE }
  \end{cases}
\]
where the $N_{\rm T}\times 2$ matrix $G_k\define\sbrac{H_i(f_k),\ H_j(f_k)}$ and $\sbrac{.}_1$ denotes the first row of the matrix.

For the evaluations, we assume a $8\times 60$ element rectangular receive antenna array at the BS and a single transmit antenna element at each UE. Since the channel measurements were done with the RX placed in 10 different rooms, we get 90 distinct combinations of $(UE_i,UE_j)$ for our interference study. For each $(UE_i,UE_j)$ location pair, we generate 300 small scale fading realizations of $H$, by varying the phase of the MPCs in~\eqn{eq:chan_tranfun_gen}. The channel capacity is then averaged over the small scale fading realizations.} Fig.~\ref{fig:cap_cdf_intf_case} plots the empirical CDF of the average capacity for different linear detectors in macrocell environment. A capacity CDF plot with no interference is also provided for reference. It can be seen that even with 480 antenna elements at the BS, the MRC detector is inferior to the MMSE detector. Because of the small elevation and azimuth spreads at the BS, there is a significant overlap in the angular power spectrum at the BS for the two UEs, thereby resulting in the degradation of the performance for MRC detector. 

\section{Conclusions}\label{sec:conclusions}
We developed a channel sounder setup consisting of a switched/virtual cylindrical antenna array at the transmitter and a switched cylindrical antenna array at the receiver and used it to carry out a 3D MIMO channel measurement campaign for an outdoor-to-indoor urban macro-cellular and urban micro-cellular environments. The increased MIMO measurement duration introduced phase drift errors in the transfer functions. Using a reference antenna measurement, we estimated and corrected for the phase drift errors in the post-processing. Using an iterative low-complexity CLEAN algorithm we extracted the MPC parameters and validated them using the map of the environment. The impact of the residual phase drift is studied using simulations, and the resulting errors in the parameter estimates and the angular spreads were quantified.

We provided the elevation and azimuth spread statistics for different UE and BS heights. The elevation spreads decreased when the BS-UE height difference is reduced. Preliminary versions of these results were provided to the 3GPP standardization group for FD-MIMO channels, and were taken into account for the establishment of their channel model. We investigated the feasibility of separating the users in different floors by elevation beamforming at the transmitter. Users were separable in $44\%$ and $54\%$ scenarios respectively for the UMa and UMi environments. Using the measured impulse responses, we studied the channel hardening aspects of Massive MIMO. Because of the small angular spreads at the BS, for the O2I environment, the MRC combiner is significantly off from the MMSE combiner. 

\appendix 
\subsection{Outlier Filtering}\label{app:outlier}
Let $h_i(f)$ be the $i^{\kth}$ snapshot of channel transfer function measurement between a given TX-RX antenna pair, measured at frequency $f$. Let $K$ denote the symmetric $10\times 10$ correlation matrix, with the ${i,j}^{\kth}$ element of the matrix corresponding to the pair-wise correlation between the snapshots $i$ and $j$. The $ij^{\kth}$ entry of the matrix is given by
\begin{equation}
K_{ij}=\abs{\int h_i(f)h_j^{\ast}(f)df}
\end{equation}

In an ideal scenario, where all the $10$ snapshots are perfectly correlated, $K$ would have been scaled all one matrix. In presence of outliers, the columns corresponding to these outlier snapshots will have significantly different entries compared to the rest. This is because, the outlier snapshot will be uncorrelated or differently correlated with all other snapshots. We use the following intuitive rule to detect and discard the outliers.

Let $g$ be a $1\times 10$ vector of column sum of matrix $K$. We use median filter to identify the indices of the outlier snapshots. The entries of the vector $g$ that are atleast $20\%$ different from the corresponding median are declared as outlier snapshots and eliminated.~\footnote{Since the number of outlier snapshots are typically few in number, the median represents the good data.}

\subsection{Other Phase Drift Estimation Techniques}
\label{app:phase_drift_est_other}
We now describe the other phase drift estimation techniques that have been tried out. For simplicity, let $H_{r,n_{\rm R}}^{\rm ref}(f_k)$ denote the channel transfer function measurement between the reference antenna and receive antenna $n_{\rm R}$, when the rotor is at position $r$. Let $h_{r,n_{\rm R}}^{\rm ref}(\tau)$ be the corresponding channel impulse response obtained by taking the inverse Fourier transform (IFFT) of the channel transfer function. 
\subsubsection{Approach 1: Combining over the RX antennas}
The transfer functions are averaged over the frequency to get the phase estimate for reach RX antenna. The phase drift estimate for a given rotor positions is obtained by maximum ratio combining (MRC) of phase estimates of the RX antenna elements.

The phase drift estimate corresponding to the rotor position $r$ is given by
\begin{equation}
\hat{\theta}_r=\angle\brac{\sum_{n_{\rm R}=1}^{N_{\rm R}}\sum_{k=1}^{N_{\rm f}}H_{r,n_{\rm R}}^{\rm ref}(f_k)}, 1\leq r\leq 60
\end{equation}
\subsubsection{Approach 2: Using the RX antenna with strongest SNR}
The transfer functions are averaged over the frequency as done earlier. For each rotor position, only the RX antenna with the strongest SNR is used to get the phase drift estimate.
\begin{equation}
\hat{\theta}_r=\angle\brac{\sum_{k=1}^{N_{\rm f}}H_{r,[\hat{1}]_r}^{\rm ref}(f_k)}, 1\leq r\leq 60
\end{equation}
where $[\hat{1}]_r$ is the index of the RX antenna with strongest SNR
\begin{equation}
[\hat{1}]_r=\arg\max_{n_{\rm R}}\sum_{k=1}^{N_{\rm f}}\abs{H_{r,n_{\rm R}}^{\rm ref}(f_k)}^2, 1\leq r\leq 60
\end{equation}
\subsubsection{Approach 3: Combining over the RX antennas, but using impulse responses}
For each rotor position, the phase drift estimate is obtained by MRC combining of the phases of the impulse response peaks over the RX antennas.
\begin{equation}
\hat{\theta}_r=\angle\brac{\sum_{n_{\rm R}=1}^{N_{\rm R}}h_{r,n_{\rm R}}^{\rm ref}\brac{\hat{\tau}^{\rm max}_{r,n_{\rm R}}}}, 1\leq r\leq 60
\end{equation}
where $\hat{\tau}^{\rm max}_{r,n_{\rm R}}$ is the location of the impulse response peak
\begin{equation}
\hat{\tau}^{\rm max}_{r,n_{\rm R}}=\arg\max_{\tau}\abs{h_{r,n_{\rm R}}^{\rm ref}(\tau)}
\end{equation}
\subsubsection{Approach 4: Using impulse response of RX antenna with strongest SNR}
For each rotor position, only the RX antenna with the strongest SNR is used to get the phase drift estimate.
\begin{equation}
\hat{\theta}_r=\angle\brac{h_{r,[\hat{1}]_r}^{\rm ref}\brac{\hat{\tau}^{\rm max}_{r,[\hat{1}]_r}}}, 1\leq r\leq 60
\end{equation}
This approach is very similar to the phase drift correction technique used in~\cite{Dunand_VTC_2007}.
\subsubsection{Performance comparison} 
We compared the performance of different phase drift correction techniques using simulations. The phase drift was modeled  as a random walk process, and the impact of residual phase drift on the angular estimates was studied. Under good SNR conditions, it has been observed that the RMSE in the angular estimates was very similar for Approach 1, 2, and 3, and comparable to the phase drift correction technique given in Sec.~\ref{sec:phase_correction}. Approach 4 gave significantly higher RMSE in the angular estimates. However, for the measurement data, the drift correction technique in Sec.~\ref{sec:phase_correction} gave better results than Approach 1, 2, and 3. The residual drift in the reference channel was significantly smaller when the drift correction technique in Sec.~\ref{sec:phase_correction} was used. This is because of the higher processing gain associated with that approach.
\bibliographystyle{ieeetr}
\bibliography{Ilmenau_Journal_draft_single_column}
\end{document}